\documentclass[runningheads]{llncs}
\usepackage{tikz}
\usetikzlibrary{arrows}
\usepackage{verbatim}
\usepackage{amsmath}
\usepackage{amssymb}
\usepackage{graphicx}
\usepackage{xypic}
\usepackage{cite}
\usepackage{array}
\usepackage{enumitem}

\usepackage[breaklinks=true]{hyperref}
\usepackage{breakcites}

\newtheorem{thm}{Theorem}
\newtheorem{lem}[thm]{Lemma}

\newcommand{\cnc}[1]{\ensuremath{\mathsf{#1}}}

\newcommand{\linseqe}{\to}
\newcommand{\braseqe}[1]{\stackrel{#1}{\prec}}
\newcommand{\brapare}[1]{\stackrel{#1}{\sim}}

\newcommand{\cpy}{\cnc{CPY}}

\newcommand{\at}[2]{\mathop{@_{#1}}{#2}}
\newcommand{\sig}{\cnc{SIG}}
\newcommand{\hsh}{\cnc{HSH}}

\newcommand{\aspn}{\cnc{ASP}}

\newcommand{\asptn}[4]{\cnc{ASP}~#1~#2~#3~#4}

\newcommand{\aspen}[2]{\cnc{U}~#1~(#2)}
\newcommand{\Ge}[2]{\cnc{G}~#1~#2}
\newcommand{\He}[1]{\cnc{H}~#1}
\newcommand{\nidt}{\cnc{n}}
\newcommand{\nid}{\emph{nonce\_id}}
\newcommand{\bits}{\emph{bits}}
\newcommand{\plc}{\emph{place\_id}}
\newcommand{\aspid}{\emph{asp\_id}}
\newcommand{\Net}[3]{\cnc{N}~{#1}~#2~(#3)}

\newcommand{\seqe}{\mathbin{;\!;}}
\newcommand{\pare}{\parallel}
\newcommand{\mt}{\xi}
\newcommand{\sign}[2]{[\![#1]\!]_{#2}}
\newcommand{\hash}[2]{\mathop{\#_{#1}}{#2}}

\newcommand{\Ke}[4]{\cnc{K}^{#1}_{#3}({#4})}
\newcommand{\Ue}[3]{\cnc{U}_{#2}({#3})}
\newcommand{\UUU}[2]{\cnc{U}_{#1}({#2})}
\newcommand{\Ne}[2]{\cnc{N}_{#1}({#2})}

\newcommand{\eval}[3]{\mathcal{E}({#1},{#2},{#3})}

\newcommand{\evalv}[3]{\mathcal{V}({#1},{#2},{#3})}

\newcommand{\bfr}[3]{{#1}\mathbin:{#2}\prec{#3}}

\newcommand{\anno}[3]{[{#1}]^{#2}_{#3}}

\newcommand{\lstp}[3]{{#1}\stackrel{#2}{\leadsto}{#3}}

\newcommand{\lstar}[1]{\stackrel{\!\!#1}{\leadsto^{{}_{*}}}}

\newcommand{\sstop}{\mathcal{D}}
\newcommand{\conf}{\mathcal{C}}

\newcommand{\terms}[0]{\cnc{\emph{t}} }

\newcommand{\evt}[0]{\cnc{\emph{E}} }
\newcommand{\evc}[0]{\cnc{\emph{e}} }

\newcommand{\pl}[0]{\cnc{\emph{P}} }

\newcommand{\bs}{\cnc{bs}}

\newcommand{\Sec}[2]{\cnc{SS} \hspace{1mm} {#1} \hspace{1mm} {#2}}
\newcommand{\Pec}[2]{\cnc{PP} \hspace{1mm} {#1} \hspace{1mm} {#2}}
\newcommand{\mtc}{\cnc{mt}}

\begin{document}
\title{An Infrastructure for Faithful Execution of Remote Attestation Protocols\thanks{This work is funded by the NSA
    Science of Security initiative contract \#H98230-18-D-0009 and
    Defense Advanced Research Project Agency contract
    \#HR0011-18-9-0001. The views and conclusions contained in this
    document are those of the authors and should not be interpreted as
    representing the official policies, either expressed or implied, of
    the U.S. Government.}}
%
%
\author{Adam Petz \and
Perry Alexander}
\authorrunning{A. Petz and P. Alexander}
%
\institute{Information and Telecommunication Technology Center \\ The
  University of Kansas \\ Lawrence, KS 66045 \\
  \email{\{ampetz,palexand\}@ku.edu}}
\maketitle              
\begin{abstract}

  Remote attestation is an emerging technology for establishing trust
  in a remote computing system.  Copland is a
  domain-specific language for specifying layered attestation
  protocols, characterizing attestation-relevant system events, and
  describing evidence bundling.  In this work we formally define and
  verify a Copland Compiler and Copland Virtual Machine for executing
  Copland protocols. The compiler translates Copland into instructions
  that are executed on the virtual machine.  The compiler and virtual
  machine are implemented as monadic, functional programs in the Coq
  proof assistant and verified with respect to the Copland event and
  evidence semantics.  In addition we introduce the Attestation
  Manager Monad as an environment for managing Copland term execution
  providing support for managing nonces, binding results of Copland
  protocols to variables, and appraising evidence results.

\end{abstract}

%
%
%

\section{Introduction}

\emph{Semantic Remote Attestation} is a technique for establishing
trust in a remote system by requesting \emph{evidence} of its
behavior, \emph{meta-evidence} describing evidence properties, and
locally \emph{appraising} the result. Remote attestation by virtual
machine introspection is introduced by Haldar and
Franz~\cite{haldar04:_seman_remot_attes} and subsequently
refined~\cite{Coker::Principles-of-R,Coker:08:Attestation:-Ev,Rowe:2016wb,Rowe:2016bi,Halling:13:Verifying-a-Pri,Challener:08:A-Practical-Gui}
to become an important emerging technology for security and trust
establishment.

In its simplest form remote attestation involves an \emph{attester} (or
\emph{target}) and an \emph{appraiser}.  The appraiser requests evidence from an attester that executes an \emph{attestation
  protocol} sequencing \emph{measurements} to gather evidence and
meta-evidence.  Upon receiving evidence from the attester, the
appraiser performs an \emph{appraisal} to determine if it can trust
the attester.

As system complexity increases so increases attestation and appraisal
complexity.  Federations of targets, systems-of-systems, privacy and
security, and layering all introduce a need for complex, multi-party
attestations.  To address this need the authors and their colleagues
developed Copland~\cite{Ramsdell:2019aa}, a language for defining and
executing attestation protocols. Copland has a formal semantics
defining measurement, where measurement is performed, measurement
ordering, and evidence bundling.

Our aspirational goal is developing a formally verified execution
environment for Copland protocols.  This work centers on a formal model for compiling and executing Copland in an
operational environment.  We define a compiler, virtual machine, and
run-time environment as functional programs in Coq, then prove them compliant with the Copland formal
semantics.  As such it informs our CakeML attestation manager
development by providing a detailed formal definition of Copland
protocol execution.


\section{Virus Checking As Attestation}

\newcommand{\dolseq}[2]{#1\linseqe#2}
\newcommand{\att}[3]{\mathop{@_{#1}}{#2}{#3}}

\newcommand{\vcpl}{p}
\newcommand{\sspl}{p}
\newcommand{\mapl}{q}
\newcommand{\maplnew}{ma}

\newcommand{\vcid}{vc}
\newcommand{\ssid}{h}
\newcommand{\osid}{os_m}

\newcommand{\vct}{t}
\newcommand{\sst}{v}
\newcommand{\ost}{os}

\newcommand{\vc}{(\asptn{\vcid}{\bar{a}}{\vcpl}{\vct})}
\newcommand{\sss}{(\asptn{\ssid}{\bar{b}}{\sspl}{\sst})}
\newcommand{\sssnew}{\att{\maplnew}{\nonce}{[\dolseq{(\asptn{\ssid}{\bar{b}}{\sspl}{\sst})}{\sig}]}}
\newcommand{\os}{(\asptn{\osid}{\bar{c}}{\vcpl}{\ost})}

\newcommand{\noncev}{n}
\newcommand{\nonce}{\{\noncev\}}

\newcommand{\aterm}{\att{\vcpl}{}{[\vc]}}
\newcommand{\bterm}{\att{\vcpl}{\nonce}{[\dolseq{\vc}{\sig}]}}
\newcommand{\btermm}{\att{\vcpl}{}{[\dolseq{\vc}{\sig}]}}
\newcommand{\cterm}{\att{\sspl}{\nonce}{[\dolseq{\sssnew}{\sig}]}}

\newcommand{\ctermnew}{\att{\vcpl}{\nonce}{[\dolseq{\sssnew}{\dolseq{\vc}{\sig}]}}}

\newcommand{\ctermm}{\att{\sspl}{}{[\dolseq{\sssnew}{\sig}]}}
\newcommand{\dterm}{\dolseq{\cterm}{\bterm}}
\newcommand{\eterm}{\att{\mapl}{\nonce}{[{\os}]}}
\newcommand{\fterm}{\dolseq{\eterm}{\dterm}}
\newcommand{\gterm}{\dolseq{\cterm}{\dolseq{\os}{\bterm}}}
\newcommand{\hterm}{\att{\mapl}{}{[\dolseq{\gterm}{\sig}]}}

\newcommand{\layered}{\att{\vcpl}{\nonce}{[\dolseq{\att{\mapl}{\nonce}{[\dolseq{(\asptn{m}{\bar{c}}{\mapl}{ss})}{\sig}]}}{\dolseq{\sssnew}{\dolseq{\vc}{\sig}}}]}}

\newcommand{\figsize}{small}


A simple motivating example for Copland is treating virus checking as
attestation.  Suppose that an appraiser would like to establish if a
target system is virus free.  The obvious approach is for the
appraiser to request virus checking results as an attestation of the
remote machine and appraise the result to determine the remote
machine's state.  The Copland phrase for this attestation is:

\vspace{-\smallskipamount}
\begin{small}
  \[\aterm\]
\end{small}
\vspace{-\medskipamount}
\vspace{-\smallskipamount}

\noindent asking platform $\vcpl$ to invoke virus checker $\vcid$ as
an attestation service provider targeting applications $\vct$ running
on $\vcpl$. 

Simply doing a remote procedure call places full trust in $\vcid$ and
its operational environment.  The target could lie about 
its results or an adversary could tamper with the virus checking
system by compromising the checker or its signature file.  An
adversary could also compromise the operational environment running
the checker or execute a man-in-the-middle replay attack.

A stronger attestation would make a request of the target that includes
an encrypted nonce to ensure measurement freshness.  The target would
decrypt the nonce, gather evidence from the checker, and return the
evidence and nonce signed using its private key.  The appraiser would
check the signature and nonce as well as checking the virus checker
results.  While the virus checker produces evidence of system state,
the signature and nonce produce \emph{meta-evidence} describing how
evidence is handled. The Copland phrase for this attestation is:

\vspace{-\smallskipamount}
\begin{small}
  \[\bterm\]
\end{small}
\vspace{-\medskipamount}
\vspace{-\smallskipamount}

\noindent adding an input nonce, $n$, and asking $\vcpl$ to sign the
measurement result.

Evidence from the virus checker may still be compromised if the virus
checker executable or signature file were compromised by an adversary.
The attestation protocol can be improved to return a measurement of
the checker's operational environment in addition to virus checking
results.  The Copland phrase for this stronger attestation is:

\vspace{-\smallskipamount}
\begin{small}
  \[\ctermnew\]
\end{small}
\vspace{-\medskipamount}
\vspace{-\smallskipamount}

\noindent where $\maplnew$ is a trusted and isolated measurement and
attestation domain with read access to $\vcpl$'s execution
environment.  $\ssid$ is a composite measurement of $\sst$, the virus
checking infrastructure--$\vcpl$'s operating system along with the
virus checking executable and signature file.  These all occur before
virus checking with the result included in a signed evidence bundle.

Measurement order is critical.  An active adversary may compromise a
component, engage in malice, and cover its tracks while avoiding
detection.  Ordering constrains the adversary by making this process
more difficult~\cite{Rowe:2016wb}.  If the virus checker is run before
its executable or signature file are hashed the adversary has much
longer to compromise the checker than if they are hashed immediately
before invoking the checker.    Ensuring measurement
order is thus critical when verifying attestation protocols and
critical to any execution or transformation of protocol
representations.

The attestation becomes yet stronger by extending to include the
signature file \emph{server} used to update signatures.  This server
operates on a different system that is remote to the system being
appraised.  However, its state impacts the overall state of the virus
checking infrastructure.  The target system can include information
about the server by performing a \emph{layered attestation} where
evidence describing the signature server is included in the target's
evidence.  The target $\vcpl$ sends an attestation request to the
server $\mapl$ that responds in the same manner as $\vcpl$:

\vspace{-\medskipamount}
\begin{scriptsize}
  \[\layered\]
\end{scriptsize}
\vspace{-\medskipamount}
\vspace{-\smallskipamount}

While the virus checking-as-attestation example is trivial, it exposes
critical characteristics of attestation protocols that motivate and
impact verification:

\begin{itemize}
  \itemsep=0pt\parskip=0pt
\item Flexible mechanism---There is no single way for performing
  attestation or appraisal.  A language-based approach for specifying
  attestation protocols is warranted~\cite{Coker::Principles-of-R}.
\item Order is important---Confidence in measurement
  ordering is critical to trusting an appraisal result.  Preserving
  measurement ordering from protocol specification to execution is a
  critical correctness property~\cite{Ramsdell:2019aa,Rowe:2016wb,Rowe:2016bi}.
\item Trust is relative---Different attestations and appraisals result
  in different levels of trust.  Formally specifying the semantics of
  attestation and appraisal is necessary for choosing the best
  protocol~\cite{Coker::Principles-of-R,Coker:08:Attestation:-Ev}.
\end{itemize}


\section{Copland Language \& Reference Semantics}

Copland is a domain specific language and formal semantics for
specifying remote attestation protocols~\cite{Ramsdell:2019aa}.  A
\emph{Copland phrase} is a term that specifies the order and
place where an attestation manager invokes attestation services.  Such
services include basic measurement, cryptographic bundling, and
remote attestation requests.  Copland is designed with expressivity
and generality as foremost goals.  As such Copland parameterizes
attestation scenarios over work leaving specifics of measurement,
cryptographic functions, and communication capabilities to protocol
negotiation and instantiation.

\subsection{Copland Phrases}
\label{sec:phrases}

The Copland grammar appears in Figure~\ref{fig:term-grammar}.  The
non-terminal $A$ represents primitive attestation actions including
measurements and evidence operations.  The constructor $\aspn$ defines
an \emph{Attestation Service Provider} and represents an atomic
measurement.  $\aspn$ has four static parameters, $m$, $\bar{a}$, $p$,
and $r$ that identify the measurement, measurement parameters, the
place where the measurement runs, and the measurement target.  A
\emph{place} parameter identifies an attestation manager environment,
and supports cross-domain measurements that chain trust across
attestation boundaries.  Parameters to an ASP are static and must be
bound during protocol selection.  Protocol participants must ensure they are properly supported by the platforms involved.

\begin{figure}
  \[\begin{array}{rcl}
      \terms&\gets&A\mid\at{p}{t}\mid(t \linseqe t)\mid
                    (t\braseqe\pi t)\mid(t\brapare\pi t) \\
      A&\gets&\asptn{m}{\bar{a}}{p}{r}
               \mid\cpy\mid\sig\mid\hsh\mid\cdots

    \end{array}\]
  \caption{Copland Phrase Grammar where: $m = \aspid\in\mathbb{N}$; $p
    = \plc\in\mathbb{N}$; $r = target\_id\in\mathbb{N}$; $\bar{a}$ is
    a list of string arguments; and $\pi = (\pi_1,\pi_2)$ is a pair of
    evidence splitting functions.
}
  \label{fig:term-grammar}
\end{figure}

Remaining primitive terms specify cryptographic operations over
evidence already collected in a protocol run.  $\cpy$, $\sig$, and
$\hsh$ copy, sign and hash evidence, respectively.  The cryptographic
implementations underlying $\sig$ and $\hsh$ are negotiated among
appraiser and target when a protocol is selected.

The key to supporting attestation of layered architectures is the
remote request operator, $\cnc{@}$, that allows attestation managers
to request attestations on behalf of each other.  The subscript $p$
specifies the place to send the attestation request and the subterm
$t$ specifies the Copland phrase to send.  As an example, the phrase
$\cnc{@_1 (@_2(t))}$ specifies that the attestation manager at place 1
should send a request to the attestation manager at place 2 to execute
the phrase \cnc{t}.  Nesting of \cnc{@} terms is arbitrary within a
phrase allowing expressive layered specifications parameterized over
the attestation environment where they execute.

The three structural Copland terms specify the order of execution and
the routing of evidence among their subterms.  The phrase
$t1 \linseqe t2$ specifies that $t1$ should finish executing strictly
before $t2$ begins with evidence from $t1$ consumed by $t2$.  The
phrase $t1\braseqe\pi t2$ specifies that $t1$ and $t2$ run in sequence
with $\pi$ specifying how input evidence is split between the
subterms. Conversely, \newpage \noindent ~$t1\brapare\pi t2$~ places no restriction on
the order of execution for its subterms allowing parallel execution.
Both branching operators produce the product of executing their
subterms.

\subsection{Concrete Evidence}
\label{sec:evidence}

Copland evidence is structured data representing the result of
executing a Copland phrase. Evidence and meta-evidence allow an
appraiser to make a trust decision about the attesting platform.  The
concrete evidence definition appears in
Figure~\ref{fig:concrete-evidence-grammar} and its structure
corresponds closely to that of Copland phrases.  Of note are the
$\cnc{mt}$ and $N$ constructors that do not correspond to a Copland
phrase.  The former stands for ``empty'', or absence of evidence, and
the latter for nonce evidence.  Raw binary data results from a
measurement and could be anything from a hash of software--the
$\cnc{bs}$ in $\aspen{\cnc{bs}}{\evc}$--to a digital signature over
evidene $e$--the $\cnc{bs}$ in $\Ge{\cnc{bs}}{e}$.  The inductive $e$
parameter accumulates sequential evidence via the $\linseqe$ phrase,
where deeper nesting implies earlier collection.  Ultimately, the
guarantee of measurement ordering comes from the Copland Virtual
Machine semantics.

\begin{figure}
  \[\begin{array}{rcl}
      \evc&\gets&\mtc\mid\aspen{\bs}{\evc}\mid
                  \Ge{\bs}{\evc}\mid\He{\bs}\mid\Net{\nidt}{\bs}{\evc}\mid\Sec{\evc}{\evc}\mid\Pec{\evc}{\evc}\mid\cdots
      
    \end{array}\]
  \caption{Conrete Evidence grammar where $\bs$ is raw binary data and
    $\nidt = \nid\in\mathbb{N}$}
  \label{fig:concrete-evidence-grammar}
\end{figure}

\subsection{Copland  LTS Semantics}
\label{sec:copland-semantics}

The Copland framework provides an abstract specification of Copland
phrase execution in the form of a small-step operational Labeled
Transition System (LTS) semantics.  States of the LTS correspond
to protocol execution states, and its inference rules
transform a Copland phrase from a protocol description to evidence.

A single step is specified as $\lstp{s_1}{\ell}{s_2}$ where $s_1$ and
$s_2$ are states and $\ell$ is a label that records
attestation-relevant system events.  The reflexive, transitive closure
of such steps, $s_1\lstar c s_2$, collects a trace, $c$, of event
labels representing a characterization of attestation activity.  $\conf(t,p,e)$ represents an
initial configuration with Copland phrase $t$, starting place $p$, and
initial evidence $e$.  $\sstop(p,e')$ represents the end of execution
at place $p$ with final evidence $e'$.  Therefore,
$\conf(t,p,e)\lstar c \sstop(p,e')$ captures the complete execution of Copland phrase $t$ that exhibits event trace $c$.

The Copland LTS semantics define a strict partial order on event
traces. The specification is
constructed inductively where: (i) Leaf nodes represent base cases and hold a single event instance; and (ii)
Before nodes ($t1\rhd t2$) and Merge nodes ($t1\bowtie t2$) are
defined inductively over terms.  Before nodes impose ordering while
Merge nodes capture parallel event interleaving where orderings within
each sub-term are maintained.  The LTS denotation function, $V$, maps
an annotated Copland term, place, and initial evidence to a corresponding Event
System.  A representative subset of this
semantics~\cite{Ramsdell:2019aa} appears in
Figure~\ref{fig:evsemantics}.

\begin{figure}
  \[\begin{array}{rcl}

  \evalv{\anno{\sig}{i}{i+1}}{p}{e}&=&
  \mathsf{SIG_{event}}(i,p,\sign{e}{p})\\

  \evalv{\anno{\asptn{m}{\bar{a}}{q}{r}}{i}{i+1}}{p}{e}&=&
  \mathsf{\aspn_{event}}(i,p,q,r,m,\bar{a},\UUU{p,q,m}{e})\\

  \evalv{\anno{\at{q}{t}}{i}{j}}{p}{e}& = &
                                                                                  \mathsf{REQ}(i,p,q,t,e)\rhd\evalv{t}{q}{e}\rhd
  \mathsf{RPY}(j-1,p,q,\eval{t}{q}{e})\\

  \evalv{\anno{t_1\brapare{\pi}t_2}{i}{j}}{p}{e}& =&
  \mathsf{SPLIT}(i,...)\rhd
  (\evalv{t_1}{p}{\pi_1(e)} \bowtie \evalv{t_2}{p}{\pi_2(e)}) \rhd
  \mathsf{JOIN}(j-1,..)\\
      
  \end{array}\]
  \caption{Event System semantics}
  \label{fig:evsemantics}
\end{figure}

Each event instance is labeled by a unique natural number and an
identifier for the place where it occurred.  Measurement and
cryptographic events correspond exactly to primitive Copland terms,
communication events $\cnc{REQ}$ and $\cnc{RPY}$ model a request and
reply interaction to a remote place, and
evidence-routing events $\cnc{SPLIT}$ and $\cnc{JOIN}$ record local
splitting and joining of evidence.  These rules are useful as a
reference semantics to characterize attestation manager execution and
denote evidence structure.  Any valid implementation of Copland
execution will obey this semantics.


\section{Copland Compiler and Virtual Machine}
\label{chap:attestation}

\newcommand{\mm}{\cnc{m1}}

\newcommand{\copcomp}{\cnc{copland\_compile}}
\newcommand{\cvmst}{\cnc{CVM\_st}}
\newcommand{\invasp}{\cnc{invoke\_ASP}}
\newcommand{\bcaxiom}{\cnc{build\_comp\_external'}}
\newcommand{\doprim}{\cnc{do\_prim}}
\newcommand{\BS}{\cnc{BS}}
\newcommand{\sttrace}{\cnc{st\_trace}}
\newcommand{\stpl}{\cnc{st\_pl}}
\newcommand{\ststore}{\cnc{st\_store}}
\newcommand{\stev}{\cnc{st\_ev}}
\newcommand{\addtrace}{\cnc{add\_tracem}}
\newcommand{\signEv}{\cnc{signEv}}
\newcommand{\sendReq}{\cnc{sendReq}}
\newcommand{\receiveResp}{\cnc{receiveResp}}
\newcommand{\doRemote}{\cnc{doRemote}}
\newcommand{\remoteEv}{\cnc{remote\_evidence}}
\newcommand{\remoteTr}{\cnc{remote\_trace}}
\newcommand{\runPar}{\cnc{runParThread}}
\newcommand{\startPars}{\cnc{startParThreads}}
\newcommand{\runcvm}{\cnc{run\_cvm}}





Copland execution is implemented as a compiler targeting a monadic,
virtual machine run-time.  The Copland
Compiler translates a Copland phrase
into a sequence of commands to be executed in the Copland Virtual
Machine (CVM).  $\copcomp$ (Figure~\ref{fig:copland-compiler-impl}) takes as input an Annotated Copland term and
returns a computation in the Copland Virtual Machine Monad.  Annotated
Copland terms extend Copland phrases with a pair of natural numbers that
represent a range of identifiers.  The compiler uses these ranges to
assign a unique label to every system event that will occur during
execution.  The LTS semantics does this similarly.  Event identifiers
play a key role in the proof that links the LTS and CVM semantics.

The Copland Virtual Machine (CVM) Monad is a state and exception monad
adapted from the Verdi framework for formally
verifying distributed
systems~\cite{Verdi-github-2016,10.1145/2737924.2737958}.
The CVM Monad implements the standard state monad primitives bind,
return, put, and get in the cannonical way.  It also provides the
standard functions for executing state monad computations
($\cnc{runState}$, $\cnc{evalState}$, $\cnc{execState}$), the
always-failing computation ($\cnc{failm}$), and getters/putters
specialized to the CVM internal state fields. Accompanying these definitions are
proofs that the CVM Monad obeys the cannonical state monad laws\cite{10.1007/978-3-642-35705-3_2}.

A general monadic computation $\cnc{St}$ takes a state parameter of
type S as input, and returns a pair of an optional return value of
type A and an updated state.  The Coq signature for $\cnc{St}$ is:

\begin{verbatim}
    Definition St(S A : Type) : Type := S -> (option A) * S
\end{verbatim}

\noindent The CVM Monad is a specialization of $\cnc{St}$ with the
$\cvmst$ type as its state structure.  $\cvmst$ is a record datatype
with fields that hold configuration data for the CVM as it
executes.

Measurement primitives build computations in the CVM Monad that perform two primary functions: simulate invocation of measurement services
and explicitly bundle the evidence results.  To simulate
measurement, $\invasp$ (Figure~\ref{fig:inv-asp}) adds a measurement event to the $\sttrace$ field of $\cvmst$, tagging it with the parameters of the service invoked along with the unique identifier $x$ derived from annotations on the originating $\aspn$ term.  Because $x$ is guaranteed unique per-protocol due to the way Copland terms are annotated, it can also serve as an abstract representation of the bit string measurement result.  This approach accounts for multiple, independent invocations of the same ASP during a protocol and captures changes in a target's state over time.  To finish, $\invasp$ bundles the result in a Copland Evidence constructor for $\aspn$s.  A single function $\doprim$ compiles all primitive Copland terms using a similar strategy.

\begin{figure}
 
\begin{verbatim}
Definition invoke_ASP (x:nat) (i:ASP_ID) (l:list Arg) : CVM EvidenceC :=
  p <- get_pl ;;  
  e <- get_ev ;;
  add_tracem [Term.umeas x p i l];;
  ret (uuc i x e).
\end{verbatim}

\caption{Example monadic measurement primitive}
\label{fig:inv-asp}
\end{figure}

When interpreting a remote request term $@_pt$ or a parallel branch
$t1\brapare\pi t2$ CVM execution relies on an external attestation
manager that is also running instances of the CVM.  To pass evidence
to and from these external components we use the shared memory
$\ststore$ component of the $\cvmst$, relying on glue code to manage
external interaction with $\ststore$.  $\sendReq$ in
Figure~\ref{fig:comm-prims} is responsible for placing initial
evidence into the shared store at index $reqi$ and initiating a request to the
appropriate platform, modeled by a
$\cnc{REQ}$ system event.  It then returns, relying on $\receiveResp$ to retrieve the
evidence result after the remote place has finished execution.
Uniqueness of event ids like $reqi$ ensures that CVM threads will not interfere
with one another when interacting with $\ststore$.

\begin{figure}
 
\begin{verbatim}
Definition sendReq (t:AnnoTerm) (q:Plc) (reqi:nat) : CVM unit :=
  p <- get_pl ;;
  e <- get_ev ;;
  put_store_at reqi e ;;
  add_tracem [REQ reqi p q (unanno t)].
\end{verbatim}

\caption{Example monadic communication primitive}
\label{fig:comm-prims}
\end{figure}

\begin{figure}
\begin{verbatim}
Definition remote_evidence (t:AnnoTerm) (p:Plc) (e:EvidenceC) : EvidenceC.

Definition remote_trace (t:AnnoTerm) (p:Plc) : list Ev.
\end{verbatim}
\caption{Primitive IO Axioms}
\label{fig:io-axioms-prim}
\end{figure}

Figure~\ref{fig:io-axioms-prim} shows two uninterpreted functions that
simulate the execution of external CVM instances.  $\remoteEv$
represents evidence collected by running the term $t$ at place $p$
with initial evidence $e$.  Similarly, $\remoteTr$ represents the list
of events generated by running term $t$ at place $p$.  There is no
evidence parameter to $\remoteTr$ because the system events generated
for a term are independent of initial evidence.  We provide
specializations of these functions for both remote and local parallel
CVM instances.  Because the core CVM semantics should be identical for
these specializations, we also provide rewrite rules to equate them.
However, their decomposition enables a straightforward translation to
a concrete implementation where differences in their glue code are
significant.

Each case of the Copland Compiler in
Figure~\ref{fig:copland-compiler-impl} uses the monadic sequence
operation to translate a Copland phrase into an instance of the CVM
Monad over unit.  The individual operations are not executed by the
compiler, but returned as a computation to be executed later.  This
approach is similar to work that uses a monadic shallow embedding in
HOL to synthesize CakeML~\cite{IJCAR18}.  The shallow embedding
style~\cite{Gill:14:DSLs-and-Synthesis} allows the protocol writer to
leverage the sequential, imperative nature of monadic notation while
also having access to a rigorous formal environment to analyze chunks
of code written in the monad.  It also leverages Coq's built-in name
binding metatheory, avoiding this notoriously difficult problem in
formal verification of deeply embedded
languages\cite{Aydemir:08:Engineering-for}.

The first three compiler cases are trivial.  The $\aspn$ term case invokes
the $\doprim$ function discussed previously that generates
actions for each primitive Copland operation.  The $\cnc{@}$ term
case invokes $\sendReq$, $\doRemote$, $\receiveResp$ in sequence.
$\sendReq$ was described previously and $\receiveResp$ is defined
similarly.  $\doRemote$ models execution of a remote CVM instance by
retrieving initial evidence from the store, adding a simulated trace of remote
events to $\sttrace$, then placing the remotely-computed evidence back
in the shared store.  Finally, the linear sequence term
$(t_1 \linseqe t_2)$ case invokes $\copcomp$ recursively on the
subterms $t_1$ and $t_2$ and appends the result in sequence.

The branch sequence case $(t_1 \braseqe\pi t_2)$ splits the initial
evidence into evidence for the two subterms using the
$\cnc{split\_evm}$ helper function.  The commands for the $t_1$ and
$t_2$ subterms are then compiled in sequence, placing initial evidence
for the respective subterm in the $\cvmst$ before executing each, and
extracting evidence results after each.  A sequential evidence
constructor combines evidence to indicate sequential execution and
emits a join event.

In the branch parallel case $(t_1\brapare\pi t_2)$ the commands for
each subterm will execute in a parallel CVM thread.  The helper
function $\startPars$ starts threads for the two subterms then appends
the trace ($\cnc{shuffled\_events}$ $el_1$ $el_2$) to $\sttrace$,
where $el_1$ and $el_2$ are event traces for the two subterms derived
from uninterpreted functions that mimic CVM execution.
$\cnc{shuffled\_events}$ is itself an uninterpreted function that
models an interleaving of the two event traces.  Event shuffling is
modeled explicitly in the LTS semantics, thus we add an axiom stating
that $\cnc{shuffled\_events}$ behaves similarly.  Similar to the
$\cnc{@}$ term case, we use the shared store to pass evidence to and
from the parallel CVM thread for each subterm.  After running both
threads, we retrieve the final evidence from the result indices,
combine evidence for the two subterms with a parallel evidence
constructor, and emit a join event.


\begin{figure}


\begin{scriptsize}
\begin{verbatim}
Fixpoint copland_compile (t:AnnoTerm): CVM unit :=
  match t with
  | aasp (n,_) a =>
      e <- do_prim n a ;;  
      put_ev e
  | aatt (reqi,rpyi) q t' =>
      sendReq t' q reqi ;;
      doRemote t' q reqi rpyi ;;
      e' <- receiveResp rpyi q ;;  
      put_ev e'
  | alseq r t1 t2 =>
      copland_compile t1 ;;
      copland_compile t2
  | abseq (x,y) (sp1,sp2) t1 t2 =>
      e <- get_ev ;;  
      p <- get_pl ;;
      (e1,e2) <- split_evm x sp1 sp2 e p ;;
      put_ev e1 ;;  copland_compile t1 ;;
      e1r <- get_ev ;;
      put_ev e2 ;;  copland_compile t2 ;;
      e2r <- get_ev ;;
      join_seq (Nat.pred y) p e1r e2r 
  | abpar (x,y) (sp1,sp2) t1 t2 =>
      e <- get_ev ;;  
      p <- get_pl ;;
      (e1,e2) <- split_evm x sp1 sp2 e p ;;
      let (loc_e1, loc_e1') := range t1 in
      let (loc_e2, loc_e2') := range t2 in
      put_store_at loc_e1 e1 ;;  
      put_store_at loc_e2 e2 ;;
      startParThreads t1 t2 p (loc_e1, loc_e1') (loc_e2, loc_e2') ;;  
      (e1r, e2r) <- get_store_at_2 (loc_e1', loc_e2') ;;
      join_par (Nat.pred y) p e1r e2r
  end.

Definition run_cvm (t:AnnoTerm) (st:cvm_st) : cvm_st :=
  execSt (copland_compile t) st.
\end{verbatim}
\end{scriptsize}
\caption{The Copland Compiler--builds computations as sequenced CVM instructions}
\label{fig:copland-compiler-impl}

\vspace{-5mm}
\end{figure}

Monadic values represent computations waiting to run.  $\runcvm$ $t$
$st$ interprets the monadic computation ($\copcomp$ $t$)
with initial state $st$, producing an updated state.  This updated
state contains the collected evidence and event trace
corresponding to execution of the input term and initial evidence. 
The evidence and event trace are sufficient to verify correctness of
$\runcvm$ with respect to the LTS semantics.



\section{Verification}
\label{sec:coq-verification}

\newcommand{\ram}{\Downarrow}
\newcommand{\cvmrec}[4]{\cnc{\{~\stev := {#1},~\ststore := {#2},~\stpl := {#3},~\sttrace := {#4}~\}} }
\newcommand{\cvmrecc}[3]{\cnc{\{~\stev := {#1},~\stpl := {#2},~\sttrace := {#3}~\}} }

\newtheorem{custthm}{Theorem}
\newenvironment{customthm}[1]{\renewcommand\thm{#1}\innercustomthm}{\endinnercustomthm}

Verifying the Copland Compiler and Copland Virtual Machine is proving
that running compiled Copland terms results in evidence and event
sequences specified by the LTS semantics.  In earlier work \cite{Ramsdell:2019aa} we proved for any
event $v$ that precedes an event $v'$ in an Event System generated by
Copland phrase $t$ ($\bfr{\evalv{t}{p}{e}}{v}{v'}$) that event also
precedes $v'$ in the trace $c$ exhibited by the LTS semantics
$\lstar {}$.  This is captured in Theorem~\ref{thm:main2}.

\begin{thm}[LTS Correctness]\label{thm:main2}
  If $\conf(t,p,e)\lstar c\sstop(p,e')$ and \\
  $\bfr{\evalv{t}{p}{e}}{v}{v'}$, then $v\ll_c v'$.
\end{thm}

To verify the compiler and virtual machine we replace the LTS
evaluation relation with executing the compiler and virtual machine
and show execution respects the same Event
System. Theorem~\ref{thm:vm_refine_thm} defines this goal:

\begin{thm}[CVM\_Respects\_Event\_System]\label{thm:vm_refine_thm}\ \\
   If $\runcvm$ ($\copcomp$ t) \\ $\cvmrecc{e}{p}{[~]}$  $\ram$ \\ $\cvmrecc{e'}{p}{c}$ and \\
  $\bfr{\evalv{t}{p}{e}}{v}{v'}$, then $v\ll_c v'$.
\end{thm}

\noindent The $\ram$ notation emphasises that
$\runcvm$ is literally a functional program written in Coq.  This
differentiates it from the $\lstar c$ notation used to represent steps
taken in the relational LTS semantics.  $\runcvm$ takes as input
parameters a sequence of commands in the CVM Monad and a $\cvmst$ structure that includes fields for initial evidence ($\stev$), starting
place ($\stpl$), initial event trace ($\sttrace$), and a shared store ($\ststore$, omitted in this theorem).   It outputs
final evidence, ending place, and a final trace.  The first
assumption of Theorem~\ref{thm:vm_refine_thm} states that running the
CVM on a list of commands compiled from the Copland phrase $t$,
initial evidence $e$, starting place $p$, and an empty starting trace
produces evidence $e'$ and trace $c$ at place $p$.  The remainder is
identical to the conclusion of Theorem~\ref{thm:main2}.

\subsection{Lemmas}

To prove Theorem~\ref{thm:vm_refine_thm}, it is enough to prove
intermediate Lemma~\ref{thm:vm_refine_lemma} that relates event traces
in the CVM semantics to those in the LTS semantics.
Lemma~\ref{thm:vm_refine_lemma} is the heart of this verification
and proves that any trace $c$ produced by the CVM semantics is
also exhibited by the LTS semantics.  We can combine
Lemma~\ref{thm:vm_refine_lemma}
transitively with Theorem~\ref{thm:main2} to prove the main correctness
result, Theorem~\ref{thm:vm_refine_thm}.
 
\begin{lem}[CVM\_Refines\_LTS\_Event\_Ordering]\label{thm:vm_refine_lemma}\ \\
  If $\runcvm$ ($\copcomp$ t) \\ $\cvmrecc{e}{p}{[~]}$  $\ram$ \\ $\cvmrecc{e'}{p}{c}$ then \\
   $\conf(t,p,e)\lstar c\sstop(p,e')$
\end{lem}

The proof of Lemma~\ref{thm:vm_refine_lemma} proceeds by induction on
the Copland phrase $t$ that is to be compiled and run through the CVM.
Each case corresponds to a constructor of the Copland
phrase grammar and begins by careful simplification and unfolding of $\runcvm$.  Each case ends with applying a semantic rule of the LTS semantics.  Lemma~\ref{thm:vm_refine_lemma} also critically proves that the CVM transforms Copland Evidence consistently with the LTS, allowing an appraiser to rely on precise cryptographic bundling and the shape of evidence produced by a valid CVM.
  
Because we cannot perform IO explicitly within Coq, we use $\sttrace$
to accumulate a trace of calls to components external
to the CVM.  This trace records every IO invocation occurring during
execution and their relative ordering.  Lemma~\ref{thm:vm_irrel}
says that $\sttrace$ is irrelevant to the
remaining fields that handle evidence explicitly during CVM execution.
This verifies that erasing the $\sttrace$ field from
$\cvmst$ is safe after analysis.

\begin{lem}[st\_trace\_irrel]\label{thm:vm_irrel}\ \\
  If $\runcvm$ ($\copcomp$ t) \\ $\cvmrec{e}{o}{p}{tr_1}$  $\ram$ \\ $\cvmrec{e'}{o'}{p'}{\_}$ and \\ \\
  $\runcvm$ ($\copcomp$ t) \\ $\cvmrec{e}{o}{p}{tr_2}$ $\ram$ \\ $\cvmrec{e''}{o''}{p''}{\_}$ then \\
  $e' = e''~and~o' = o''~and~p' = p''$
\end{lem}

A key property we would like to be true of the CVM semantics is that
event traces are \emph{cumulative}.  This means that existing event
traces in $\sttrace$ remain unmodified as CVM execution
proceeds.  Lemma~\ref{thm:foo} encodes this, saying:  If a CVM state
with initial trace $m~++~k$ interprets a compiled Copland term $t$ and
transforms the state to some new state $st'$, and similarly $t$
transforms a starting state with initial trace $k$ (the suffix of the
other initial trace) to another state $st''$, then the $\sttrace$
field of $st'$ is the same as $m$ appended to the $\sttrace$ field of
$st''$.  This ``distributive property'' over traces is vital in several
other Lemmas that simplify event insertion and trace composition.

\begin{lem}[st\_trace\_cumul]\label{thm:foo}\ \\
  If $\runcvm$ ($\copcomp$ t) \\ $\cvmrec{e}{o}{p}{m~\cnc{++}~k}$ $\ram$ $st'$  and \\ \\
  $\runcvm$ ($\copcomp$ t) \\ $\cvmrec{e}{o}{p}{k}$ $\ram$ $st''$ then \\
  ($\sttrace~st')~=~m~\cnc{++}~(\sttrace~st'')$
\end{lem}

\newcommand{\wf}{\cnc{well\_formed}}

\subsection{Automation}

There are many built-in ways to simplify and expand expressions in
Coq. Unfortunately, it is easy to expand either too far or not enough.
The Coq $\cnc{cbv}$ (call-by-value) tactic unfolds and expands as much
as it can, often blowing up recursive expressions making them
unintelligible.  The milder $\cnc{cbn}$ (call-by-name) tactic often
avoids this, but fails to unfold user-defined wrapper functions.  For
this reason, we define custom automation in Ltac.  First we define a
custom ``unfolder'' that carefully expands everything from primitive monadic operations like
bind and return, to CVM-specific helper functions like $\invasp$.

Next we define a larger simplifier that uses the $\cnc{cbn}$ tactic to
conservatively simplify expressions having concrete arguments.  We
then repeatedly invoke the custom unfolder followed by $\cnc{cbn}$ and
other conservative simplifications.  This custom simplification is the
first step in most proofs and is repeated throughout as helper
Lemmas transform the proof state to expose more reducible expressions.

\begin{lem}[abpar\_store\_non\_overlap]\label{thm:store_acc}\ \\
  If $\wf$ $(abpar~\_~\_~t_1~t_2)$ and \\
  range $t_1$ = (a,b) and \\
  range $t_2$ = (c,d) then \\
  a $\cnc{\ne}$ c and b $\cnc{\ne}$ c and b $\cnc{\ne}$ d
\end{lem}

A final custom automation involves Lemmas that ensure accesses to the
shared store do not overlap when interpreting Copland terms that
interact with external components.  When compiling the branch parallel
term we derive indices from term annotations and use them to insert initial evidence
and retrieve final evidence from the store.  We must prove arithmetic
properties like Lemma~\ref{thm:store_acc} to show that store accesses
do not overlap.  The proof follows from the definition of the $\wf$
predicate and the annotation strategy.  We provide Ltac scripts to
recognize proof states that are blocked by store accesses within
larger Lemmas and discharge them using Lemma~\ref{thm:store_acc}.


\section{Attestation Manager(AM) Monad}
\label{sec:am-monad}

While the CVM Monad supports faithful execution of an individual
Copland phrase, many actions before and after execution are more
naturally expressed at a layer above Copland.  Actions preceding
execution prepare initial evidence, collect evidence results from
earlier runs, and generate fresh nonces.  Actions following CVM
execution include appraisal and preparing additional Copland phrases
for execution.  These pre- and post- actions are encoded as statements
in the Attestation Manager (AM) Monad.

The AM Monad is so named because it manages multiple
executions of Copland phrases and appraises
resulting evidence.  The $\cnc{run\_avm(t,n)}$ command runs an
entire Copland phrase $t$ with initial evidence $n$ inside the CVM
Monad, lifting its evidence result into the AM Monad.  This is
accomplished with the Copland Compiler and VM.  By invoking
$\cnc{run\_avm}$, the AM Monad 
does not perform measurements directly, but rather relies on a
well-defined interface to the CVM allowing the AM to abstract away
details of
Copland phrase execution.

The AM Monad demands a computational context with a combination of
stateful, immutable environment, error, and IO functionality.  State
is required for remembering nonces and evidence results of Copland
phrases for use as initial evidence and during appraisal.  A read-only
environment can hold configuration data needed during appraisal such
as public keys, handles to local appraisal routines, and expected
measurement values.  While these values are configurable by the
platform owner, they should remain immutable during appraisal
execution.  Error functionality allows graceful handling of
communication failure and ill-formed evidence structures in a Copland
response.  Finally, IO supports running CVM computations, along with
invoking black-box appraisal primitives and generating nonces.  An early
prototype of the AM Monad in Haskell~\cite{haskell-am-github-2020}
uses monad transformers to compose the State monad with Reader, Error,
and IO to implement necessary computational effects.  An initial formal definition of the AM Monad in Coq, including nonce management and Copland phrase invocation, is complete.  The design of appraisal and its verification are ongoing.

\subsection{Nonce Management}

Using nonces is a common mechanism for preventing a man-in-the-middle
adversary from re-transmitting stale measurements that do not reflect
the current state of the system.  Nonces are critical to attestation
and appear in Copland as initial evidence passed
alongside the Copland phrase in an attestation request.  Since
evidence collection is cumulative in the CVM semantics, executing a
Copland phrase builds up evidence around the nonce embedded as initial
evidence.

Nonces are a unique form of evidence because they do not
have a corresponding Copland phrase.  Instead, they are generated and
stored in the AM Monad state, passed as initial evidence, then
retrieved during appraisal.  To manage multiple outstanding nonces, we
added two fields to the state called $\cnc{am\_nonceId}$ and
$\cnc{am\_nonceMap}$.  The $\cnc{am\_nonceId}$ field is a natural number
that represents the unique next nonce id.  It is initialized to 0,
and incrememted upon each subsequent nonce generation within a single
AM Monad computation.  $\cnc{am\_nonceMap}$ is the mapping of nonce
ids to their raw binary values, used to remember the nonce for
comparison during appraisal.

\subsection{Appraisal}
\label{sec:appraisal}

Appraisal is the final step in a remote attestation protocol where an indirect observer of a target platform must analyze evidence in order to determine the target's trustworthiness.  Each appraiser has its own standards, and thus two appraisers may make a different decision given the same evidence.  Regardless of its level of scrutiny, an appraiser must have a precise understanding of the structure of evidence it examines.  The Copland framework provides such a shared evidence structure, and Copland phrases executed by the CVM produce evidence with a predictable shape.  In addition to the knowledge of evidence structure, an appraiser's configuration must include appropriate cryptographic keys and ``golden'' measurement values before it can unbundle and perform semantic checks over the evidence.  The AM Monad provides an ideal context to perform appraisal because it can access golden measurement and nonce values, and also link evidence to the Copland phrase that generated it.  This combination of capabilities enables automatic synthesis of appraisal routines left for future work.


\section{Related Work}

Integrity measurement tools include both static \cite{tboot, Sailer:04:Design-and-impl} and dynamic
\cite{8599862, 10.1007/978-3-642-30244-2_13,
10.1145/1133058.1133063, Shi:05:Bind:-A-fine-gr,
8228638, 10.1007/978-3-642-15497-3_12} approaches that
support both baseline and recurring measurements of target systems.  Higher-level measurement frameworks support userspace monitoring \cite{pendergrass2017maat, 8511973, Lauer2018UserCenteredAF, Petz:2019aa}, kernel-level introspection \cite{Loscocco:07:Linux-kernel-in}, and attestation of embedded/IOT platforms \cite{10.1145/2592798.2592824, 10.1145/2744769.2744922, Wedaj:2019:DDA:3328797.3325822}.  The framework presented in the current work is designed as a common operational environment for such tools, plugging them in as ASPs and composing their measurement results as Copland Evidence.

Prior work in analysis of remote attestation systems involves
virtualized environments \cite{Coker::Principles-of-R,
Berger:06:vTPM:-Virtualiz, Lauer2018UserCenteredAF}, comparing
protocol alternatives \cite{Rowe:2016wb, Rowe:2016bi}, and
semantics of attestation \cite{Datta:09:A-logic-of-secu,
Francillon:2014aa}.  These analyses articulate the complexities
in the attestation design space and lay a foundation for future
frameworks.  Coker et. al is of particular influence, as the design
principle of Trustworthy Mechanism was a primary motivation of this
work. 

HYDRA \cite{10.1145/3098243.3098261} (Hybrid Design for
Remote Attestation) was the first
hardware/software hybrid RA architecture to build upon
formally-verified components, and that achieved design goals laid
out in their prior work \cite{Francillon:2014aa}.  ERASMUS
\cite{DBLP:journals/corr/CarpentRT17} levereged HYDRA as a base
security architecture, but added real-time assurances for
resource-critical embedded platforms.  VRASED (Verifiable Remote Attestation for Simple Embedded Devices)
extended these ideas to a concrete RA design, becoming the first formal
verification ``of a HW/SW co-design implementation of any security
service'' \cite{10.5555/3361338.3361437}.  They specify end-to-end
security and soundness properties in LTL that define necessary and sufficient properties to achieve secure RA.  Their approach to verification--extracting core properties of attestation functionality from Verilog specifications of hardware, then manually incorporating
independent verification of cryptographic software--is
comparable with the design goals of the Copland effort.  Although
their end-to-end security guarantees are complete and convincing, our
attestation managers support a wide range of
attestation scenarios on diverse platforms, rather than on a fixed,
embedded platform.


\section{Future Work and Conclusion}
\label{sec:conclusion}

In this work we have verified the Coq implementation of a Copland
compiler and monadic virtual machine.  Specifically, we proved that
the output of compilation and virtual machine execution respects the
small-step LTS Copland semantics.  Artifacts associated with this
verification are publicly available on github
\cite{copland-avm-github-2020-nfm21}.  All proofs are fully automated and
the only admitted theorems are axioms that model interaction with IO
components external to the core virtual machine.


\begin{figure}[hbtp]
  \centering
  \begin{tikzpicture}[->,>=stealth',shorten >=1pt,auto,node distance=1.5cm,
  thick,main node/.style={rectangle,
    font=\sffamily,minimum height=7mm,minimum width=10mm}]

  \node[main node] [color=gray!50] (Request) {$R$};
  \node[main node] [color=gray!50] (Proposal) [below of=Request] {$\langle P\rangle $};
  \node[main node] (Phrase) [below of=Proposal] {Protocol};
  \node[main node] (CAM) [below of=Phrase] {CVM};
  \node[main node] [color=gray!50] (CakeML) [below of=CAM] {CakeML/seL4};
  \node[main node] [color=gray!50] (Binary) [below of=CakeML] {Binary};
  
  \node[main node] [color=gray!50] (Meas) [node distance=3.0cm, right of=Binary] {Evidence};
  \node[main node] (Bits) [color=gray!50,node distance=3.0cm, right of=CakeML] {Evidence};
  \node[main node] (VME) [node distance=3.0cm, right of=CAM] {Evidence};
  \node[main node] (Evidence) [node distance=3.0cm, right of=Phrase] {Evidence Shape};
  \node[main node] [color=gray!50] (EvidenceVec) [node distance=3.0cm, right of=Proposal] {$\langle E\rangle$};
  \node[main node] [color=gray!50] (Result) [node distance=3.0cm, right of=Request]
  {$(E,\preceq,\top,\bot)$};
  \node[main node] (IN) [color=gray!50,node distance=2.0cm, left of=Request] {Request};
  \node[main node] (OUT) [color=gray!50,node distance=2.0cm, right of=Result] {Result};

  \path[every node/.style={font=\sffamily\small, fill=white,inner sep=1pt}]
    (IN) edge [color=gray!50] (Request)
    (Request) edge [color=gray!50] node[left=1mm] {Negotiation} (Proposal)
    (Proposal) edge [color=gray!50] node[left=1mm] {Selection} (Phrase)
    (Phrase) edge node[left=1mm] {copland\_compile} (CAM)
    (CAM) edge [color=gray!50] node[left=1mm] {Synthesis} (CakeML)
    (CakeML) edge [color=gray!50] node[left=1mm] {CakeML Compile} (Binary)
    (Binary) edge [color=gray!50] node[below=1mm] {} (Meas)
    (CakeML) edge [dashed,color=gray!50] node[below=1mm] {} (Bits)
    (CAM) edge node[above=1mm] {run\_cvm} (VME)
    (Phrase) edge [dashed] node[below=1mm] {} (Evidence)
    (Proposal) edge [dashed,color=gray!50] node[below=1mm] {} (EvidenceVec)
    (Request) edge [dashed,color=gray!50] node[below=1mm] {} (Result)
    (Meas) edge [color=gray!50] node[right=1mm] {} (Bits)
    (Bits) edge [color=gray!50] node[right=1mm] {} (VME)
    (VME) edge node[right=1mm] {} (Evidence)
    (Evidence) edge [color=gray!50] node[right=1mm] {Abstraction} (EvidenceVec)
    (EvidenceVec) edge [color=gray!50] node[right=1mm] {Appraisal} (Result)
    (Result) edge [color=gray!50] (OUT)
    ;
\end{tikzpicture}

  \caption[Verification Figure]{Verification stack showing
    verification dependencies and execution path. Solid lines
    represent implementations while dashed lines represent
    mathematical definitions.}
  \label{fig:certification-fig}
\end{figure}
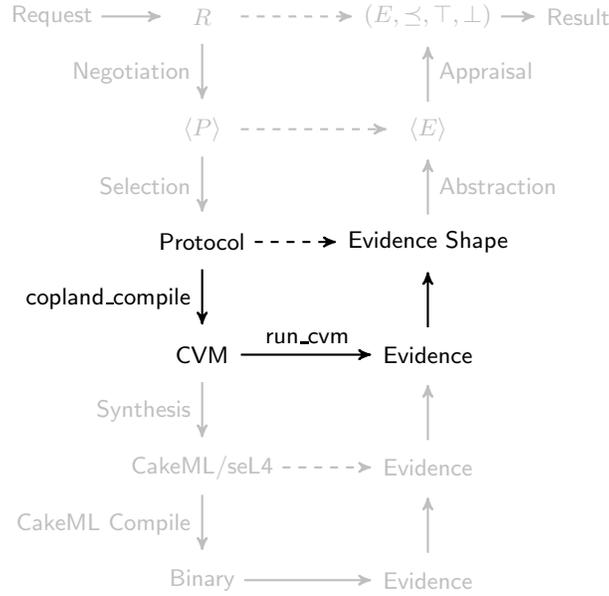

\newpage
Verification of the compiler and vm are part of our larger effort to
construct a formally verified attestation system.  Our verification
stack in Figure~\ref{fig:certification-fig} shows this work in context
with gray elements that represent supporting work or work in
progress.  Above protocol execution is a negotiation process that
selects a protocol suitable to both appraiser and target.  Ongoing
work will formally define a ``best'' protocol and verify the
negotiated protocol is sufficient and respects privacy policy of all
parties.

Below protocol execution is an implementation of the Copland Compiler
and Copland Virtual Machine in
CakeML~\cite{Kumar:2014:CVI:2535838.2535841} running on the verified
seL4
microkernal~\cite{Klein:10:seL4:-formal-ve,Klein:09:seL4:-formal-ve}.
CakeML provides a verified compilation path from an ML subset to
various run-time architectures while seL4 provides separation
guarantees necessary for trusted measurement.  We are embedding the
semantics of CakeML in Coq that will in turn be used to verify the
compiler and vm implementations.  Unverified implementations of both
components have been implemented and demonstrated as a part of a
hardened UAV flight control system.

When completed our environment will provide a fully verified tool
stack that accepts an attestation request, returns evidence
associated with that request, and supports sound appraisal of that
evidence.  This work is an important first step
creating a verified operational environment for attestation.


\newpage

\nocite{Coker::Principles-of-R}

%
%
%
\bibliographystyle{splncs04}
\bibliography{nfm21}
\end{document}